\documentclass{an}
\usepackage{graphicx}
\usepackage{times}
\Volume{01}              
\Year{2004}              
\Month{04}               
\Pagespan{1}{7}      	 

\begin{document}
\lhead[\thepage]{A.N. Karata\c{s} et al.: Analysis of $UBV$ Photometry in Selected Area 133}
\rhead[Astron. Nachr./AN~{\bf XXX} (200X) X]{\thepage}
\headnote{Astron. Nachr./AN {\bf 32X} (200X) X, XXX--XXX}

\title{Analysis of $UBV$ Photometry in Selected Area 133}

\author{Y.~Karata\c{s} 
\and  S. Bilir
\and  S. Karaali
\and  S. G. Ak}
\institute{Istanbul University Science Faculty, 
           Department of Astronomy and Space Sciences, 
           34119, University-Istanbul, Turkey}
\date{} 

\abstract{We interpret the de-reddened $UBV$ data for the field SA 133 to 
deduce the stellar density and metallicity distributions function. The 
logarithmic local space density for giants, $D^{*}(0)=6.40$, and the  
agreement of the luminosity function for dwarfs and sub-giants with the one 
of Hipparcos confirms the empirical method used for their separation. The 
metallicity distribution for dwarfs gives a narrow peak at $[Fe/H]=+0.13$ 
dex, due to apparently bright limiting magnitude, $V_{o}=16.5$, whereas 
late-type giants extending up to $z \sim4.5$ kpc from the galactic plane 
have a  multimodal distribution. The metallicity distribution for giants 
gives a steep gradient $d[Fe/H]/dz=-0.75$ dex kpc$^{-1}$ for thin disk and 
thick disk whereas a smaller value for the halo, i.e. $d[Fe/H]/dz=-0.45$ 
dex kpc$^{-1}$.
}

\correspondence{sbilir@istanbul.edu.tr}

\maketitle

\section{Introduction}
Star count analyses of Galactic structure have provided a picture 
of the basic structural and stellar populations of the Galaxy. Examples and 
reviews of these analyses can be found in Bahcall (1986), Gilmore \& Reid (1983),
Majewski (1993), and very recently in Robin, Reyl\'e, \& Cr\'ez\'e, (2000),
Robin et al. (2003), Chen et al. (2001), and Siegel et al. (2002). Most of these 
programs have been based on photographic surveys; the Basel Halo Program 
(Becker 1965) has presented the largest systematic photometric survey of the 
Galaxy (Fenkart 1989a-d; Del Rio \& Fenkart 1987; Fenkart \& Karaali 1987;1990). 
The Basel Halo Program photometry is currently being recalibrated and reanalysed, 
using an improved calibration of the $RGU$ photometric system which comprises 
homogeneous magnitudes and colours of about 20000 stars in a total of fourteen 
fields, distributed along the Galactic meridian through the Galactic centre 
and the Sun. While the ensemble of the full survey data are being analysed 
by comprehensive modeling of the density, luminosity, and metallicity 
distributions of the different Galactic population 
components (Buser, Rong, \& Karaali (BRK) 1998; 1999), the data in each individual 
field are being used as well for a detailed study of the luminosity function 
(Ak, Karaali, \& Buser 1998; Karata\c{s}, Karaali, \& Buser 2001) and 
metallicity gradient (Rong, Buser, \& Karaali 2001; Karaali et al. 2003; Karaali, 
Bilir, \& Buser 2004). In this paper we present the investigation of an individual 
field in the $UBV$ photometry, as some other works carried out as a part of Basel 
Halo Program, due to its Galactic position ($l=6^{o}.5$). The abundance data have 
been derived from broad band photometric estimates, primarily using the 
UV-excess (Sandage 1969). The technique of UV-excess to derive metal abundances 
has been applied for the photographic data in some works (e.g. Gilmore, Wyse, 
\& Jones 1995; Karaali et al. 2004) though it broadens the metallicity distribution. 

Section 2 details the general characteristics of the photometric data set and the 
method. Section 3 discusses the density function with two galactic models and 
comparison of the resulting luminosity function with that of Hipparcos (Jahreiss \& 
Wielen (JW) 1997) and Gliese \& Jahreiss (GJ 1992). In Section 4 we discuss the 
metallicity distribution. Section 5 provides a conclusion.  

\section{Data and Method}
\subsection{De-reddening of Photometry}
The $UBV$ data of 1729 stars in the field SA 133 ($l=6^{\circ}.5$, $b=+10^{\circ}.3$) 
were taken from the Basel Photometric Catalogue No. VIII (Becker et al. 1982) and the 
distances of 137 stars, brighter than $V=15$ mag, to the standard $(U-B)_{o}-
(B-V)_{o}$ main sequence along the reddening line (Fig. 1) are used for reddening 
estimation. Among these stars 122 have only photographical $UBV$ data whereas the 
remaining 15 stars have photoelectrical $UBV$ data, taken from Mermilliod \& 
Mermilliod (1994), additional to their photographical ones. The mean colour excess 
for all of these stars is $E(B-V)=0.25$ mag and their standard error for the mean,
$\sigma=0.01$ mag. This value is consistent with $E(B-V)$ = 0.33 and 0.37 mag 
derived for our field by Schlegel, Finkbeiner, \& Davis (1998), and Burnstein 
\& Heiles (1982), respectively. However $E(B-V)=0.18$ mag, given by Fenkart et al. 
(1986) is less than our value. Thus, corrections applied to $U-B$ and $V$ due to 
reddening are $E(U-B)= 0.72E(B-V) = 0.18$ and $A_ {V}=3.0 E(B-V)= 0.75$, 
respectively. The procedure applied here for de-reddening of the $UBV$ data is 
the same as the one applied for all fields investigated in the Basel Halo Program 
(Hersperger 1973; Becker \& Svopoulos; 1976 Becker \& Fang 1982; Becker 
\& Hassan 1982).

The catalogue error given by Fenkart et al. (1986) for $U$, $B$, and $V$ magnitudes, 
for the magnitudes brighter than 17, is $\pm$ 0.02 mag which corresponds to $\pm$ 
0.03 mag in $B-V$ and $U-B$. The colour magnitude diagram, $V_{o}-(B-V)_{o}$, 
in Fig. 2 indicates a limiting magnitude of $V_{o}=16.5$ mag for our field.  

\begin{figure}
\resizebox{7cm}{7cm}{\includegraphics*{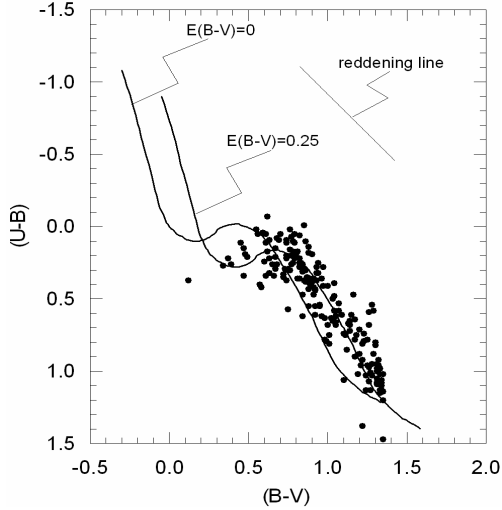}} 
\caption {Two-colour diagram for stars used for estimation of the colour-excess, $E(B-V)=0.25$ mag.}
\end {figure}

\begin{figure}
\resizebox{7cm}{7cm}{\includegraphics*{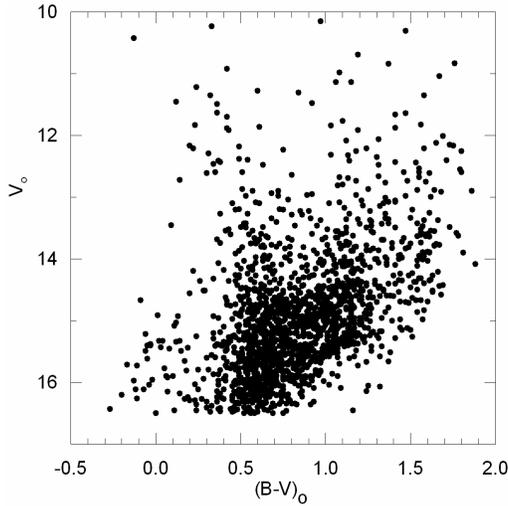}} 
\caption {De-reddened colour magnitude diagram for the field SA 133.}
\end {figure}

\subsection{Separation of dwarfs and evolved stars}

The $V$-fractioned two-colour diagrams in Fig. 3 show that there is a considerable 
amount of metal-rich stars consistent with the direction of our field. The 
positions of some stars in our two-colour diagrams seem to be affected by the 
scattering which is unavoidable in the photographic photometry for faint stars. 
The most conspicuous appearance of these stars are those occupying the position 
of the most metal-rich stars though they are not so large in number to affect 
our results. Contrary to the metal-rich stars, extreme metal-poor stars 
($[Fe/H] < -3.0$ dex) are sparse, especially in apparently bright $V-$ intervals. 
Such stars exist in all two-colour diagrams in the $UBV$ and $RGU$ 
systems and sometimes they are in an unexpected considerable amount of number 
(Karata\c{s} et al. 2001). Hence, they are excluded from statistics without regarding 
their identity which can be binary stars, extra-galactic objects or their position 
may be affected by relatively large photographic errors which is the case for faint 
magnitudes. The iso-metallicity lines, for the range $-3\leq[Fe/H]\leq+1$ dex, 
in Fig. 3 were adopted from Lejeune, Cuisinier, \& Buser (1997).

The separation of dwarfs and evolved stars (sub-giants or giants) was carried out to 
obtain a luminosity function consistent with the local luminosity function of 
nearby stars due to Gliese \& Jahreiss (1992) and Jahreiss \& Wielen (1997). 
The procedure of this separation is based on the fact that the local 
luminosity functions obtained for many fields indicated a systematic excess of star 
counts relative to the luminosity function of nearby stars for the fainter segment, 
i.e. $M(G)\geq6$ mag, and a deficit for the brighter segment, $M(G)<5$ mag.
The excess cited is due to the contamination of evolved stars and it is difficult to 
separate them from the dwarfs due to the lack of a spectral band sensitive to the 
surface gravity, $log~g$, in the $UBV$ and $RGU$ photometric systems. However, it 
was shown in some recent works, if apparently bright stars with $M(G)\geq6$ mag on 
the main sequence are removed to the category of evolved stars, both segments cited 
above converge to the local luminosity function of nearby stars (Karaali 1992; 
Ak et al. 1998; Karata\c{s} et al. 2001; Karaali et al. 2004). 
Because this process causes decrease in the number of absolutely faint stars, 
$M(G)\geq6$ mag, whereas it increases the number of bright stars. The apparent 
limiting  magnitude of bright stars considered, lies within $15 < G <16$ mag. 
The best fit between the local luminosity function deduced in this work and the 
local luminosity function of nearby stars resulted, from iterations, when stars 
apparently brighter than $V_{o}=15.5$ mag and absolutely fainter than $M(V)= 5.5$ 
mag on the main sequence were assumed to be evolved stars. Thus, brighter 
absolute magnitudes were attributed to them and they were separated to giant - 
or sub-giant - category according to their magnitudes, i.e. $M(V)\leq2$ mag and 
$M(V)>2$ mag, respectively.

\begin{figure}
\resizebox{8.4cm}{20.4cm}{\includegraphics*{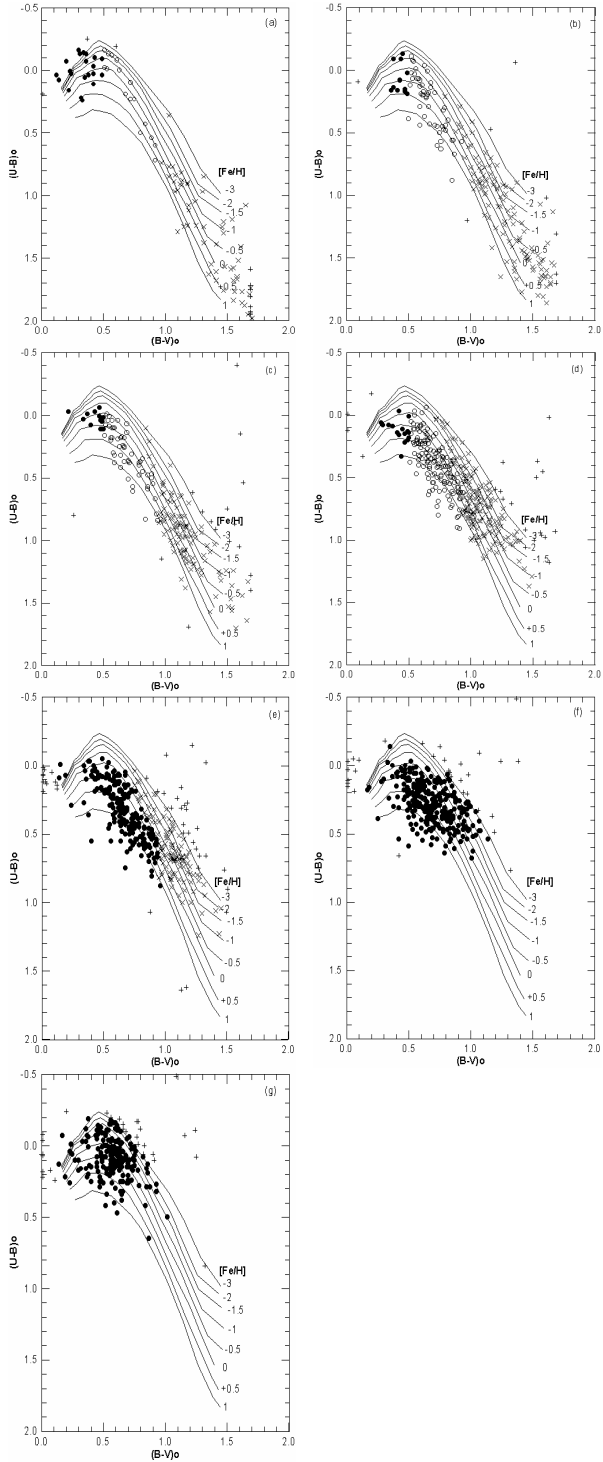}} 




\caption {The $V$-fractioned two-colour diagrams: (a) $V_{0}\leq13.0$, 
(b) (13.0 14.0], (c) (14.0 14.5], (d) (14.5 15.0], (e) (15.0 15.5], 
(f) (15.5 16.0], and (g) (16.0 16.5]. Symbols: ($\bullet$): dwarfs, 
(o): sub-giants, (x): giants, and $(+)$: not included into statistics 
due to its position in the two-colour diagram. The iso-metallicity 
lines, $-3\leq[Fe/H]\leq+1$ dex, were adopted from the synthetic 
data of Lejuene et al. (1997).}   
\end{figure}

\subsection{Absolute magnitude and metallicity determination}

The absolute magnitudes and metallicities are evaluated by means of two methods. 
For dwarfs with $[Fe/H]\geq-1.75$ dex, we adopted the metallicity-luminosity 
calibration of Laird, Carney, \& Latham (LCL 1988) which is given in terms of the offset 
in absolute $V$ magnitude from the Hyades main sequence, for which they obtained

\begin{equation}
M(V)_{(Hyades)}=5.64(B-V)+1.11 
\end{equation}
The metallicity-dependent offset from this fiducial sequence that LCL derived 
($\Delta {M^{H}_{V}}$) is, for a star of given $(B-V)$ colour and given UV-excess 
$\delta$;

\begin{equation}
^{\Delta {M^{H}_{V}}=[\frac{2.31-1.04{(B-V)_{o}}}{1.594}][-0.6888\delta+53.14\delta^2-97.004\delta^3]}
\end{equation}
LCL state this calibration to be valid for $\delta ~\leq 0.25$, which corresponds 
to $[Fe/H]\geq-1.75$ dex with Carney (1979) transformation of $\delta$ into 
$[Fe/H]$, as used by LCL and recently by Gilmore et al. (1995).

\begin{equation}
$[Fe/H]$ = 0.11 - 2.90\delta - 18.68~\delta^{2}  
\end{equation}
As already cited, the calibration of LCL is valid only for dwarfs with 
$[Fe/H]\geq-1.75$ dex. Hence another method was necessary for additional 
metal-poor dwarfs and evolved stars of any metallicity, for absolute magnitude 
and metallicity derivation. With a optimistic approach, the $(U-B, B-V)$ 
two-colour diagram calibrated for iso-metallicity lines 
($-3\leq[Fe/H]\leq+1$ dex) for dwarfs and evolved stars individually, by 
means of the synthetic data of Lejeune et al. (1997), can provide 
metallicities with appropriate interpolation. For absolute magnitude derivation 
of giant stars, we calibrate $M(V)-(B-V)_{0}$ relation, using published data 
from of various Galactic globular clusters to cover a large range of metal 
abundances. The globular clusters used are M92 ($[Fe/H] = -2.24$ dex), 
M5 (-1.40), 47 Tuc. (-0.71), and M67 (0.00), respectively, (Stetson \& Harris 
1988; Richer \& Fahlman 1987; Hesser et al. 1987; Montgomery, Marschall, 
\& Janes 1993). 

For the metallicity calibration of Carney (1979), an error of $\pm$ 0.02 mag 
in $\delta$ yields an uncertainty of 0.3 dex. The accuracy of the calibration 
of the Basel Library Spectra (Lejeune et al. 1997) is in level of less than 
0.05 mag, implying an uncertainty in derived $[Fe/H]$ of up to 0.35 dex. The
uncertainties in derived metallicities from two method are approximately the 
same. We assumed that the same holds for late-type giants. The uncertainty 
$\pm$ 0.03 mag in $(B-V)_{o}$, cited above, corresponds to $\pm$ 0.2 mag in 
absolute magnitude $M(V)$ derived by the colour-magnitude diagrams of four 
clusters. Combination of the uncertainities in $V_{o}$ and $M(V)$ in the 
equation gives the distance $r$ to the star, i.e.

\begin{equation}
V_{o} - M(V) = 5log (r) - 5 
\end{equation}
amounts to an uncertainty of 10$\%$, in the distance estimate, which is less than 
the one (20$\%$) cited by Gilmore et al. (1995).

\begin{table*}

\caption[]{The logarithmic space densities $D^{*}$ for dwarfs and sub-giants 
for all population types (distances in $kpc$, volumes in $pc^{3}$. The symbols 
are defined in the text. Underlines indicate limiting distance of completeness.}

\begin{center}
\begin{tabular}{ccccccccccccccccccc}
\hline
\multicolumn{2} {c} {M(V)} & \multicolumn{1} {c} {$\rightarrow$} &   \multicolumn{2} {c} {(2 3]} &\multicolumn{2} {c}
{(3 4]} & \multicolumn{2} {c} {(4 5]} & \multicolumn{2} {c} {(5 6]} & \multicolumn{2} {c} {(6 7]} &
\multicolumn{2} {c} {(7 8]}& \\
\hline
$r_{1}$ - $r_{2}$ &  $\Delta V_{1,2}$ &  $ \bar{r}$   & N & D* &  N & D* &  N & D* & N & D* & N &  D* & N & D*\\
\hline
0.00 - 0.40 & 1.22~(03) & 0.32 & & & & & & & &      &            &            &          2 &  7.22          \\
0.00 - 0.63 & 4.85~(03) & 0.50 & & & & & & & &      &         24 &    {7.69} &            &                 \\
0.00 - 1.00 & 1.93~(04) & 0.79 &  8 & 6.62& 35 & 7.26 &  &  &  55 & 7.45 &  &  &      &                     \\
0.00 - 1.58 & 7.68~(04) & 1.26 &     &    &   &  &  83 & 7.03 &     &      &         &       &       & \\
0.40 - 0.63 & 3.63~(03) & 0.54 &     &            &       &  & &  & & & & &  30 & 7.92 \\ \cline{14-15}
0.63 - 1.00 & 1.44~(04) & 0.86 &     &    & &  &  & &  &  & 60 &   7.62 &    &         \\ \cline{12-13}
1.00 - 1.58 & 5.75~(04) & 1.36 & 15  & 6.42  &  46 & 6.90 &  & & 74 & 7.11 & 7 & 6.09 &  &   \\\cline{10-11}
1.58 - 2.51 & 2.29~(05) & 2.15 & 52  & 6.36  & 176 & 6.89 & 114 & 6.70 & 9 & 5.59 & & & &  \\ \cline{6-7} \cline{8-9}
2.51 - 3.98 & 9.12~(05) & 3.41 & 39  & 5.63  &  76 & 5.92 &  10 & 5.04 & & & & & &         \\ \cline{4-5}
3.98 - 6.31 & 3.63~(06) & 5.40 & 40  & 5.04  &   4 & 4.04 & &  &  &  &  & &    &           \\
6.31 - 10.00& 1.44~(07) & 8.55 &  3  & 3.32  & & &  & & & & &  &     &                     \\
\hline
 & & $s (\pm)$& & 0.27 & & 0.13 &  &0.13&  &  0.17  &  &  0.00  & & 0.53  \\
\hline
\end{tabular}  
\end{center}
\end{table*}

\section{Density and luminosity functions}

Logarithmic space densities, $D^{*}=log D + 10$ are evaluated for dwarfs and 
sub-giants for consecutive absolute magnitude intervals, i.e. (2 3], 
(3 4], (4 5], (5 6], (6 7], and (7 8] (Table 1), and for late-type giants 
(Table 2), where $D(r)= N/\Delta V_{1,2}$, N: number of stars in the partial volume 
$\Delta V_{1,2}=(\pi/180)^{2}(\sq/3)(r_{2}^{3}-r_{1}^{3})$, $r_{1}$, $r_{2}$: 
limiting distances of $\Delta V_{1,2}$, $\sq$: 0.19 square-degree, apparent size 
of the field, $\bar{r}=[(r^{3}_{1}+r^{3}_{2})/2]^{1/3}$ centroid distance of the 
corresponding partial volume $\Delta V_{1,2}$.

\begin{table}
\caption[]{The logarithmic space densities for late-type giants 
(symbols as in Table 1). The standard deviation is $s$ = $\pm$0.39.} 
\begin {center}
\begin{tabular}{cccc}
\hline
$r_{1}-r_{2}$ & $\Delta V_{1,2}$ & $ \bar{r}$ & N~~~~D* \\

\hline
 0.00-3.98 &   1.22 (6) &       3.16 &        111~~~~5.96 \\
 3.98-6.31 &   3.63 (6) &       5.40 &        112~~~~5.49 \\
 6.31-10.0 &   1.44 (7) &       8.55 &        122~~~~4.93 \\
10.00-15.85 &  5.75 (7) &      13.55 &        108~~~~4.27 \\
15.85-25.12 &  2.29 (8) &      21.48 &         58~~~~~3.40 \\
25.12-39.81 &  9.12 (8) &      34.05 &         40~~~~~2.64 \\
\hline
\end{tabular}
\end{center}
\end{table}

The density functions are compared with the galactic model of Gilmore \& Wyse 
(GW, 1985), and Buser et al. (BRK, 1998; 1999) given in the form 
$\Delta logD(r)=log D(r,l,b)-logD(0,l,b)$ versus $r$, where $\Delta log D(r)$ is 
the logarithmic difference of the densities at distances $r$ and at the Sun. 
Thus $\Delta log D(r)=0$ points the logarithmic space density at $r=0$ which is 
available for luminosity function determination. The comparison is carried out 
as explained in several studies of the Basel fields (Del Rio \& Fenkart 1987; 
Fenkart \& Karaali 1987), i.e. by shifting the model curve perpendicular to the 
distance axis until the best fit to the histogram results at the centroid 
distances (Fig. 4a-g). Both models involve two disk components, thin and thick 
disk, and the de Vaucauleurs spheroid. The parameters for two models are given 
in Table 3. Although the two set of parameters (especially the scale heights) 
are different from each other, the model gradients (solid line for the model GW 
and dashed line for the model BRK) are close to each other in Fig. 4a-f, 
probably due to relatively short distances ($r<4$ kpc). Actually, the last 
histogram section for giants (Fig. 4g), $25.12<r\leq39.81$ kpc, show that the 
model gradient for BRK begins to diverge from the model gradient for GW at 
larger distances. There is adequate agreement between GW and BRK models and the 
observed density functions for dwarfs and sub-giants within the limiting 
distances of completeness marked by horizontal thick lines in Table 1 and the 
same agreement results, when one includes the local luminosity function. The 
luminosity function close to the Sun: $\varphi^{*}(M)$, i.e. the logarithmic 
space density for the stars with $M$ $\pm~0^{m}.5$ at $r$ = 0 is the $D^{*}$ 
-value corresponding to the intersection of the model-curve with the ordinate 
axis of the histogram concerned. Fig. 5 shows the agreement between the 
luminosity function resulting from comparison of our space density data with 
the GW model and the local luminosity functions of GJ and JW.

\begin{table}
\caption{Parameters for the models of GW and BRK. The symbols give: $n_{i}$: 
local space density, $H_{i}$: scale-height, $h_{i}$: scale-length, $c/a$: axis 
ratio for the halo (scale heights and scale lengths in pc).}
\center
\begin{tabular}{cccc}
\hline
Population &  Parameter    &         GW &        BRK \\
\hline
 thin disk & $n_{1}$       &          1 &          1 \\
           & $H_{1}$       &        300 &        290 \\
           & $h_{1}$       &       4000 &       4010 \\
\hline
thick disk & $n_{2}/n_{1}$ &       0.02 &       0.06 \\
           & $H_{2}$       &       1000 &        910 \\
           & $h_{2}$       &       4000 &       4250 \\
\hline
      halo & $n_{3}/n_{1}$ &      0.001 &     0.0005 \\
           & $c/a$         &       0.85 &       0.84 \\
\hline
\end{tabular}  
\end{table}

\begin{figure}
\resizebox{8cm}{14cm}{\includegraphics*{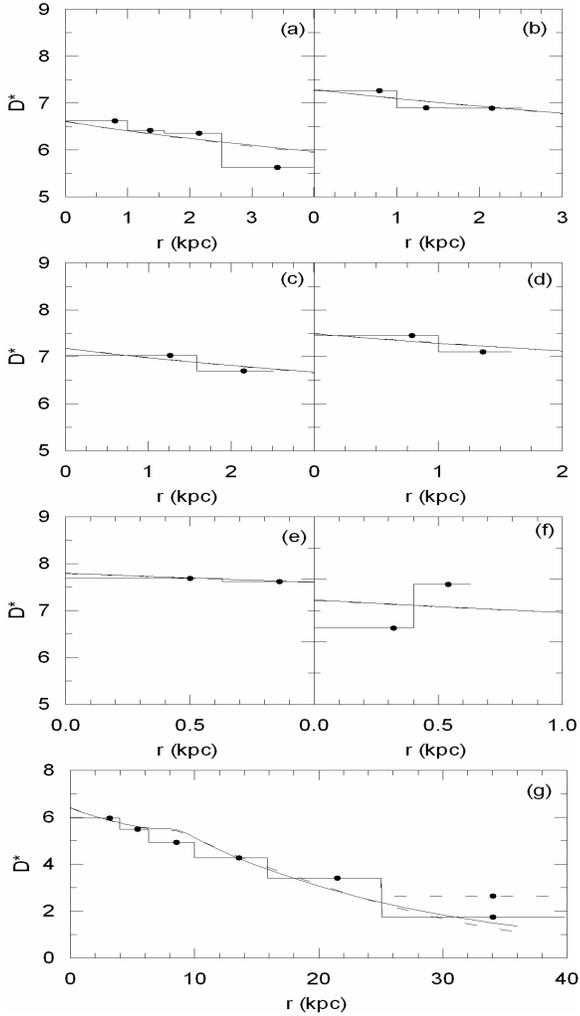}} 
\caption {Logarithmic space density functions for dwarfs and sub-giants for 
six consecutive absolute magnitude intervals: (a) (2 3], (b) (3 4], 
(c) (4 5], (d) (5 6], (e) (6 7], (f) (7 8]; and for late-type giants (g). 
Thin solid and dashed lines represent model gradient of GW and BRK, 
respectively. The last histogram section ($25.12<r\leq39.81$ kpc) with 
dashed line corresponds to the density for stars in excess.}
\end {figure}

However there is some difference between GW model and the observed density 
function for late type giants (Fig. 4g), extending up to $r\approx40$ kpc 
($z\approx7$ kpc distance to the Galactic plane). Actually there are 35 stars 
in excess in the last distance interval which cannot be adopted as dwarfs 
due to the agreement of the luminosity function with the ones cited above. 
It is interesting that the same stars do show an excess in the comparision 
between BRK model and the observed density function. Although the gradients 
for two models do not fit in every distance interval, i.e. the gradient 
for the model of BRK is slight above the model gradient of GW at $r\approx16$ 
kpc and it is below for the distance interval $r\geq26$ kpc, they present the 
same local space density. We do not have enough information for the exact 
identification of these stars in excess. They may be related to the fact that 
a number of stars have been eliminated because they have been assigned a too 
faint metallicity ($[Fe/H]<-3$ dex) or they may be extra-galactic objects. If 
the stars mentioned in the first case were not eliminated the observed 
densities in the preceding distance intervals would be increased and there 
would not be appeared stars in excess in the last distance interval. In this case 
the local space density would be closer the the one of Gliese (1969). The second 
case seems also a solution for the problematic 35 stars, because the number of 
extra-galactic objects increases when one goes to faint magnitudes. 
Unfortunately the identification  of the extra-galactic objects in this field 
could not be carried out by the University of Minnesota due to its low Galactic 
latitude ($b=+10^{\circ}.3$). If we omit these stars in excess, we obtain a 
logarithmic local space density $D^{*}(0)=6.40$ between those of Gliese (1969) 
and Fenkart (1989c), i.e. $\odot = 6.64$ and $D^{*}(0)=6.29$, respectively. The 
errors for the density functions are given in Tables 1 and 2, and in Fig. 5 in 
standard deviations.

\begin{figure}
\resizebox{8cm}{5cm}{\includegraphics*{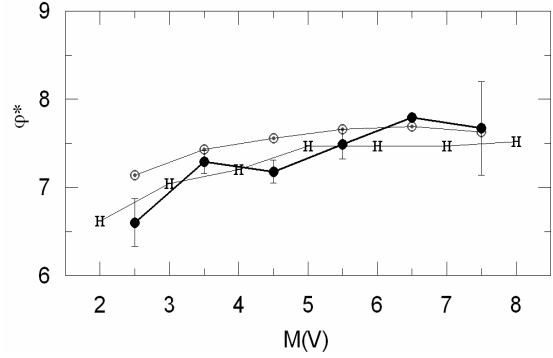}} 
\caption {Luminosity function for SA 133 resulting from 
comparison of observed histograms with the model gradient 
GW ($\bullet$), and confronted to the luminosity 
functions of GJ ($\odot$), and JW (H).}
\end {figure}

\section{Metallicity distribution}

A gaussian fit to the metallicity distribution for dwarfs and sub-giants 
(Fig. 6a) shows a high and narrow peak at $[Fe/H]=+0.13$ dex and a long 
metal-poor tail, though with less contribution to the total metallicity, 
expected for a low-latitude field to the galactic centre direction. 
Contrary to dwarfs and sub-giants, metallicity distribution for late-type 
giants which extend up to $r=25$ kpc relative to the Sun or $z=4.5$ kpc 
above the galactic plane thus forming a sub-sample of different population 
types is multi-modal (Fig. 6b). A gaussian fit to Fig. 6b gives two peaks 
at $[Fe/H]=+0.12$, $-0.83$ dex, and additionally a flat metal-poor tail 
extending down to $[Fe/H]=-3$ dex. 

\begin{figure}
\resizebox{8cm}{9cm}{\includegraphics*{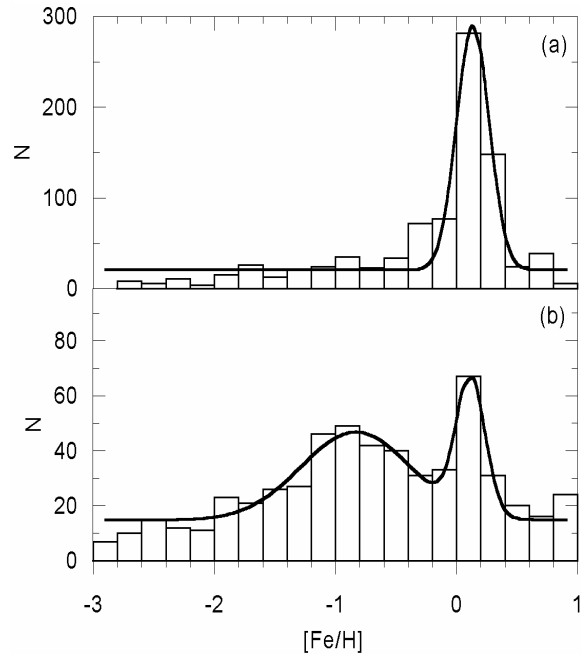}} 
\caption[]{Metallicity distribution for dwarfs and sub-giants (a), 
and late-type giants (b). Curves in (a) and (b) are the fitted 
gaussian distributions.}
\end{figure} 

Each histogram in Fig 7a-c is fitted a gaussian curve with a mean 
$<[Fe/H]>=-0.14$, $-0.78$, and $-1.10$ dex. $z$-heights shown in 
the panels are $<z> = 0.96$, $1.72$, and $2.20$ kpc, respectively. One can 
notice that a systematic shift from metal rich stars to the metal poor ones 
in Fig. 7a-c, where the mean metallicity as a function of $z$ distance is displayed. 

Comparison of the mean metal abundances and $<z>$ distances with any of 
these panels gives metallicity gradients as follows\\
{\scriptsize
$d[Fe/H]/dz=[(-0.78)-(-0.14)]/(1.72-0.96)=-0.84$~dex~kpc$^{-1}$ \\
$d[Fe/H]/dz=[(-1.10)-(-0.14)]/(2.20-0.96)=-0.77$~dex~kpc$^{-1}$ \\
$d[Fe/H]/dz=[(-1.10)-(-0.78)]/(2.20-1.72)=-0.67$~dex~kpc$^{-1}$ \\
}

The gaussian of Fig. 7d shows a peak at $<[Fe/H]>=-1.35$ dex, with a 
$<z>=3$ kpc, which reveal three metallicity gradients close to each 
other but different than cited above, when compared with the data in 
each former panels. The mean of these metallicity gradients, i.e. 
$d[Fe/H]/dz=-0.45$ dex kpc$^{-1}$, give the indication that metallicity 
gradient is less steep in the outward of the Galaxy.

\begin{figure}
\resizebox{6cm}{15cm}{\includegraphics*{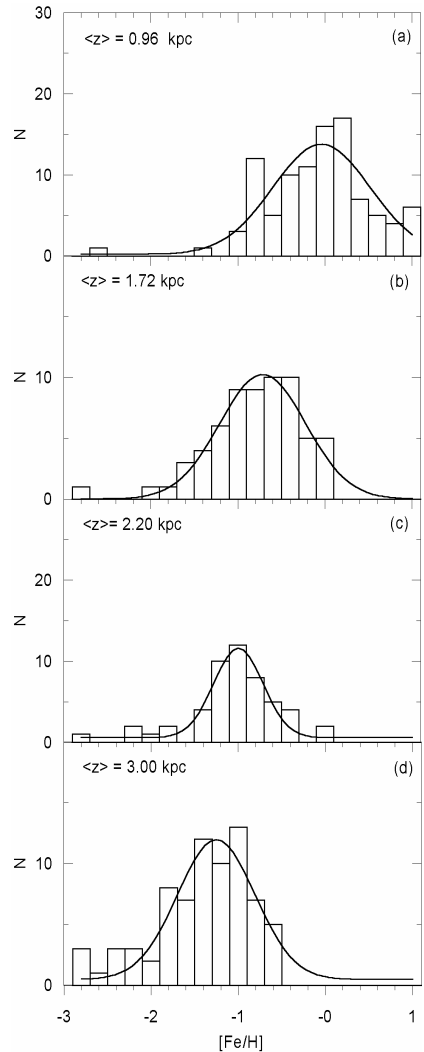}} 
\caption[] {Metallicity distributions for late-type giants for four 
$z$-intervals which reveal a metallicity gradient into the direction 
of the Galactic centre. Curves in each panel are the fitted gaussian 
distributions.} 
\end {figure}

The mean $z$-distance 1 kpc corresponding to a mean distance 5.7 kpc in the 
Galactic plane lies in the dominant region of thin disk, and the corresponding 
metallicity gradient, $d[Fe/H]/dz$ = -0.84 dex kpc$^{-1}$, is close to the one given by 
Trefzger, Pel, \& Gabi (1995) ($d[Fe/H]/dz$ = -0.55 $\pm$ 0.10 dex kpc$^{-1}$) up to 
$z$ = 0.9 kpc for a field investigated by means of Walraven $VBLUW$ photometry. These 
authors claim a less steep gradient, -0.23 $\pm$ 0.04 dex kpc$^{-1}$, for $z<4$ kpc, 
however. The metallicity gradient for thin disk, deduced from the investigation of seven fields 
by Rong et al. (2001), very recently, is $d[Fe/H]/dz > -0.6$ dex kpc$^{-1}$, not contradicting 
with our results, within the limits of accuracy. Stars in the third and fourth panels 
($<z>$ = 2.2 and 3 kpc) in our work sample the galactic halo for which metallicity 
gradient is not controversial. Thus, the metallicity gradient for stars in the second 
panel ($<z>$ = 1.72 kpc) to be discussed, for the region occupied by these stars is 
dominant by thick disk. Such a metallicity gradient contradicts some formation histories 
postulated for the formation of the (classical) thick disk. Until recent years this component
of our Galaxy was assumed to have a mean metal abundance $[Fe/H]\sim-0.60$ dex, 
a scale-height $H\leq 1$ kpc, and its space number density 2 - 8 \% of the thin disk
in the solar neighbourhood. Additionally, and more important, it was argued that
the stars of thick disk were formed from a merger into the Galaxy 
(cf. Norris 1996 and references within), a formation mechanism unlikely to leave an
abundance gradient. Some recent works suggested that the thick disk is more important 
component of the Galaxy, extending up to $\sim 3-5$ kpc from the galactic plane 
(Majewski 1993) with a metal-poor (Norris 1996) and a metal-rich tail 
(Carney 2000; Karaali et al. 2000; Karaali et al. 2004). Hence a revision of the 
formation scenario of the thick disk may be required. Furthermore, the works of 
Reid \& Majewski (1993) and Chiba \& Yoshii (1998) also suggest a metallicity 
gradient for the thick disk.  

\section{Conclusion}

The agreement of the luminosity function resulting from the comparison of GW model 
with the logarithmic space density functions for stars with $2<M(V)\leq8$ mag, 
and the logarithmic space density function for late-type giants with the same model 
(and also with the model BRK) confirm the separation of dwarfs and evolved stars 
in our field. 

There is a concentration at $[Fe/H]=+0.13$ dex and a long tail down to 
$[Fe/H]\sim -3$ dex for the metallicity distribution for dwarfs and sub-giants as 
expected for a low latitude field into the galactic centre direction with a short 
apparent limiting magnitude, i.e. $V_{o}$=16.5 mag. Whereas late-type giants which 
lie up to large distances relative to the Sun or up to large $z$-heights from the 
galactic plane have multi-modal metallicity distribution. However, the weighted mean 
metal-abundances of late-type giants in our sample is $<[Fe/H]>=-0.7$ dex, rather 
close to the one claimed by Morrison \& Harding (1993), i.e. $-0.8$ dex, 
who investigated only the K giants in two square degree field almost symmetric to 
ours relative to the galactic plane ($l=350^{\circ}$, $b=-10^{\circ}$) and 
to Harding's (1996) finding, $<[Fe/H]>=-0.8$ dex, for G and K giants. 

In this work, we observed metallicity gradient in each of the populations. 
This is different than the one usually expected: the metallicity gradient 
may appear when there is a mixing of populations for which the relative 
proportion of the populations changes along the line of sight. In such case 
it results in a gradient which may not be present in each of the populations. 
However, in the present data there is a variable proportion of the populations 
when $z$ is increasing, indicating a gradient in the populations. Additionally, 
it is rather steep in the regions dominated by thin and thick disks, i.e. 
$d[Fe/H]/dz= -0.75$ dex kpc$^{-1}$. As cited in Section 4, Reid \& Majewski 
(1993) and Chiba \& Yoshii (1998) also suggest a metallicity gradient for the 
thick disk. All these works would have implications on the discussion about the 
scenario of formation of the thick disk. However, we recognize that there are 
significant statistical uncertainties in our resuls, due to photographic data. 
It is also probable that the direction of our field ($l=6^{\circ}.5$) may play 
a specific role on our results.

\begin{acknowledgements}
This work was supported by the Research Fund of the University of Istanbul. 
Project number: 1039/031297. Also, we would like to thank to Professor E. 
Hamzao\u glu for reading the whole manuscript and correction, and to the referee 
A. Robin for her comments. 
\end{acknowledgements}

\end{document}